Shaping Nonlinearity in Reset Controllers for Precision Motion Systems

Nima Karbasizadeh[1], S. Hassan HosseinNia[1]

[1] Department of Precision and Microsystem Engineering, Delft University of Technology, Delft, the Netherlands

s.h.hosseinniakani@tudelft.nl

**Abstract**

The precision motion industry has an ever-increasing demand for faster, more precise and more robust controllers. From the perspective of frequency domain and loopshaping technique, this demand has pushed the linear controllers to their inherent limits, namely, the waterbed effect and Bode's phase-gain relationship. Mathematically, complex-order transfer functions are not bound by Bode's phase-gain relationship. However, implementing them in practice is a challenge to be solved. This extended abstract will show review the previous work of the authors in shaping nonlinearities in reset controllers and contribute and propose an overall architecture for reset control systems to approximate complex order controllers.

## 1. Complex-order transfer functions

A derivative of complex order can be defined in a variety of ways. But it is commonly indicated by the operator $D^{\alpha+j\beta}$, where $\alpha + j\beta \in \mathbb{C}$. The simplest corresponding transfer function in the Laplace domain will be $G(s) = s^{\alpha+j\beta}$. The frequency response of such transfer function is given in [1] as

$$20\log_{10}|G(j\omega)| = 20\alpha \log_{10}\omega + 20\log_{10} e^{-\frac{\beta\pi}{2}}$$
$$\angle G(j\omega) = \frac{\alpha\pi}{2} + \beta \log(10)\log_{10}\omega$$
(1.1)

When $\alpha < 0$ and $\beta > 0$, the frequency response will show a negative gain slope and a positive phase slope, for which there is no practical implementation method in the linear domain. However, such frequency response is highly desirable, especially in precision motion control, since one can, for example, increase the bandwidth of the system without sacrificing the phase margin [2]. However, this complex-order transfer is not implementable directly. Approximations with linear transfer functions through methods like CRONE often result in unstable poles or non-minimum phase zeros appearing in the controller [1]. Hence, recently researchers turned to nonlinear controllers for such approximation.

## 2. Reset control systems

Reset control systems are one of the simplest nonlinear control systems used to approximate complex-order transfer functions. Reset controllers were first proposed by Clegg [3] in the form of a reset integrator which can improve the transient response of a control system. In order to address the drawbacks and exploit the benefits, the idea was later extended to more sophisticated elements such as "FirstOrder Reset Element" [4, 5] and "Second-Order Reset Element" [6] or using Clegg's integrator in the form of PI+CI [7] or resetting the state to a fraction of its current value, known as partial resetting [8]. However, most of the research done in reset control theory lack either frequency domain and loopshaping perspective or industrial applicability.

Among different forms of reset control systems, the following is the most common

$$\Sigma_R = \begin{cases} \dot{x}_r(t) = A_r x_r(t) + B_r e(t), & \text{if } e(t) \neq 0 \\ x_r\left(t^+\right) = A_\rho x_r(t), & \text{if } e(t) = 0 \\ u(t) = C_r x_r(t) + D_r e(t) \end{cases} \quad (2.1)$$

where $A_r, B_r, C_r, D_r$ denote the state space matrices of the Base Linear System (BLS) and reset matrix is denoted by $A_\rho = \text{diag}(\gamma_1, ..., \gamma_n)$ which contains the reset coefficients for each state. $e(t)$ and $u(t)$ represent the input and output for the reset controller, respectively.

## 3. Reset control systems approximating complex-order transfer functions

Recently some steps have been taken to use reset control in approximating complex-order transfer functions [1, 9]. One specific type of reset controller named "Constant in Gain, Lead in Phase" (CgLp), which is introduced in [9], has a constant gain over a tunable frequency range while providing phase lead. According to (1.1), such frequency domain behaviour resembles that of $s^{j\beta}$. Such an element can be created by multiplying a reset lag element and a linear lead element. Figure 1 depicts the frequency response of such an element. This element's advantageous phase lead is due to the fact that reset elements have less phase lag than their linear counterparts. The frequency response of this element does not follow Bode's phase-gain relationship. This element can partly replace the derivative action in a PID framework.

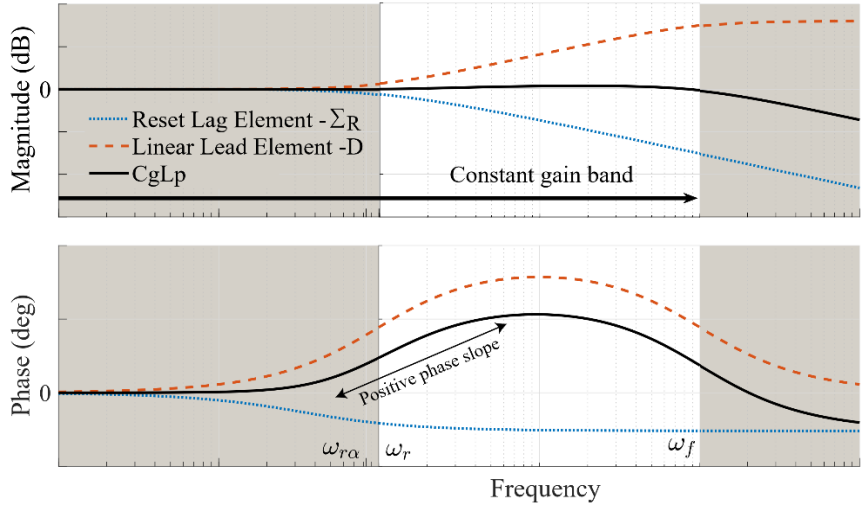

Figure 1: The concept of combining a reset lag and a linear lead element to form a CgLp element. The figure is from [9].

However, this figure shows the approximate frequency response behaviour of CgLp based on Describing Function (DF) method. In this method, only the first harmonic of the Fourier transform of the response is considered, and higher-order harmonics are neglected. In some cases, this assumption can be inaccurate regarding severe performance deterioration caused by higher-order harmonics [10].

## 4. Shaping nonlinearities in reset control systems

The higher-order harmonics directly result from nonlinearity in the reset control systems. Therefore, they cannot be totally eliminated since the resultant system will be linear and thus bound by linear systems' inherent limitations. Nevertheless, the nonlinearity and, therefore, the higher-order harmonics in reset control systems can be shaped to minimise the disadvantages and maximise the advantages. This section will briefly introduce some methods for shaping the nonlinearity of reset control systems.

### 4.1. Shaping nonlinearity by phase shaping the reset condition

In [11, 12], two new architectures for reset controllers are introduced. They both rely on the same concept for shaping nonlinearities in reset controllers. These studies use the following definition for the reset control system

$$\Sigma_R = \begin{cases} \dot{x}_r(t) = A_r x_r(t) + B_r e(t), & \text{if } x_{rl}(t) \neq 0 \\ x_r(t^+) = A_\rho x_r(t), & \text{if } x_{rl}(t) = 0 \\ u(t) = C_r x_r(t) + D_r e(t) \end{cases} \quad (4.1)$$

It should be noted that the reset condition is not the zero-crossing of the input to the reset element but a different signal named $x_{rl}(t)$. These studies show that assuming a sinusoidal input to the element, $\psi(\omega) := \angle \dfrac{X_r(j\omega)}{X_{rl}(j\omega)}$ changes the behaviour of the element, where $X_r, X_{rl}$ stand for Laplace transform of $x_r(t)$ and $x_{rl}(t)$. In the special case of $\psi(\omega)=0$, $x_{rl}(t)=0$ implies $x_r(t)=0$. Then it implies that at the reset instant, $x_r(t^+)=x_r(t)=0$. It means that the after-reset state value is the same as the current state value, and practically, the reset action has no effect, and the system always remains in the linear regime; thus, higher-order harmonics will be zero at all frequencies satisfying $\psi(\omega)=0$. These researches also study $\psi(\omega)\neq 0$ and how it affects the behaviour of the system. The shaping of the phase $\psi(\omega)$ is done through a shaping filter. Figure 2 shows the block diagram of the suggested architecture of [12].

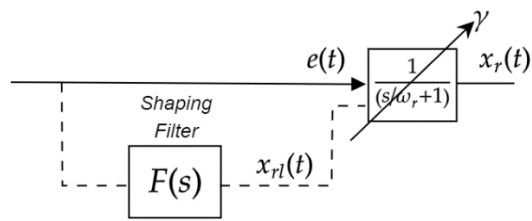

Figure 2: The architecture used in [12] for shaping the phase $\psi(\omega)$. Arrow indicates the resetting action.

In [12], a method of tuning for shaping filter, $F(s)$, is introduced, which allows different phase slopes to be achieved. In other words, with a combination of this element and a linear lead filter, an approximation of $s^{j\beta}$ for a desired value of $\beta$ can be achieved. Figure 3 shows the frequency response of a sample linear PID controller compared to the first harmonic of a sample Complex-Order Controller (CLOC) designed based on the approach of [12]. Although both controllers provide the same phase advantage at the intended cross-over frequency of 100 Hz, CLOC shows a higher gain at lower frequencies. Figure 4 shows the closed-loop sensitivity of the same controllers when controlling a mass plant.

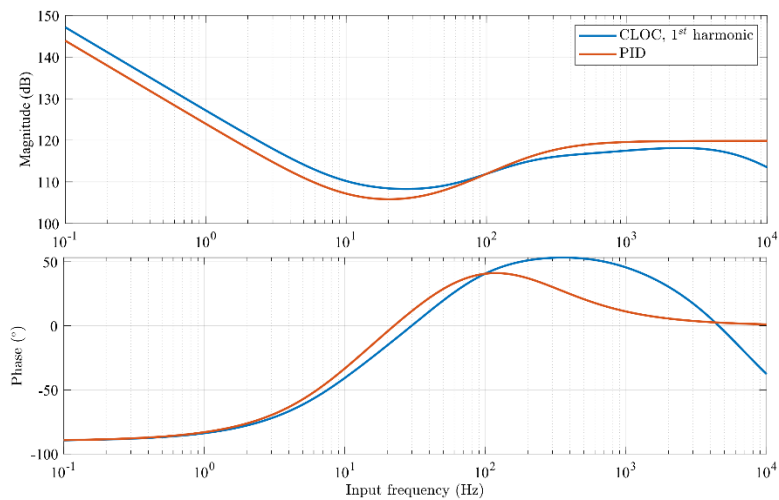

Figure 3: FRF of a linear PID vs. DF of a CLOC.

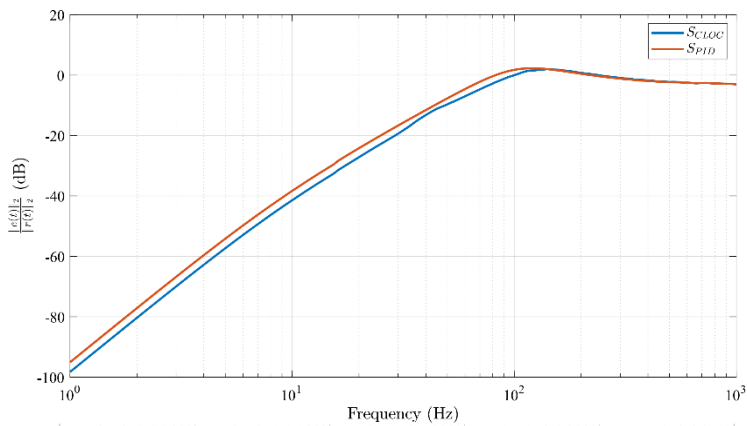

Figure 4: Closed-loop sensitivity function of a PID and a CLOC controlling a mass plant.

In [11], the focus of shaping $\psi(s)$ is to confine the nonlinearity of the reset control system to a desired range of frequencies. In other words, the higher-order harmonics can be band-passed using this method. The suggested architecture is shown in Figure 5.

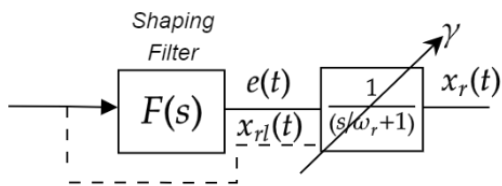

Figure 5: The architecture used in [11] for shaping the phase $\psi(\omega)$. Arrow indicates the resetting action.

The tuning method for shaping filter, $F(s)$ to achieve such an objective is introduced in [11]. Figure 6 compares a band-passed CgLp element with a conventional one in terms of the magnitude of higher-order harmonics. One can notice that higher-order harmonics are significantly reduced outside the desired range of [1 10] rad/s. One can also notice that the phase advantage is now also limited to this frequency range. However, phase advantage in precision motion control applications is usually required around the cross-over frequency region to provide stability and robustness and improve the transient response. Providing phase lead in lower frequencies also increases the higher-order harmonics and deteriorates the steady-state precision in terms of tracking and disturbance rejection. In Figure 6, the frequencies for which $\psi(\omega) = 0$ are indicated by $\omega_{lb}$. One can notice that higher-order harmonics' magnitude goes to zero at these frequencies, indicating linear behaviour at these frequencies.

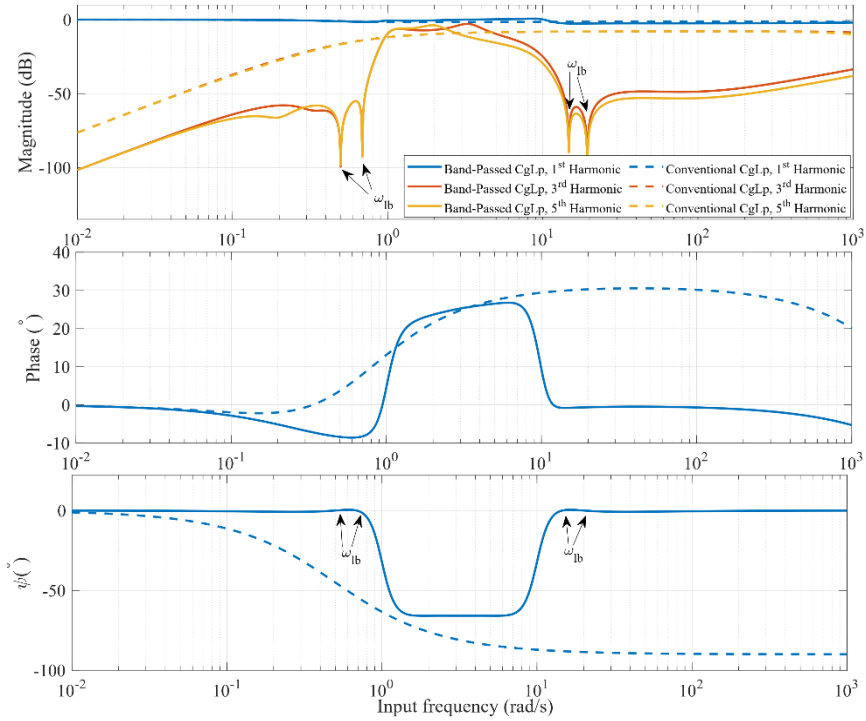

Figure 6: Comparison of a band-passed CgLp and a conventional one.

### 4.2. Shaping nonlinearity using a continuous reset structure

Another important common property of all reset elements in the literature is the discontinuity of the output signal. This property is a cause for the presence of high-frequency content in the signals and other practical issues such as big jumps in control signals that can saturate actuators. In [13], a new architecture, as shown in Figure 7, is introduced, which can make the output of the reset controller continuous.

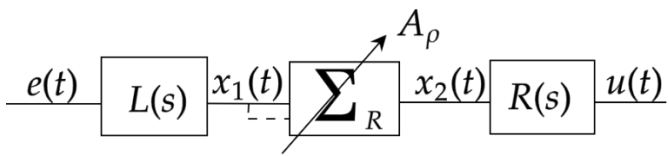

Figure 7: The architecture used in [13] for shaping nonlinearity in reset controllers.

In this architecture $L(s) = \dfrac{s/\omega_l + 1}{s/\omega_h + 1}$ and $R(s) = \dfrac{1}{s/\omega_l + 1}$. By taking $\omega_l < \omega_h$ and a large enough $\omega_h$ one can see that in the linear domain $R(s)$ and $L(s)$ would cancel each other. However, this is not the case for nonlinear controllers such as reset controllers.

In the closed loop, assuming $e(t)$ is the closed-loop error because of the lead behaviour of $L(s)$, the system will reset not only based on the error but also based on its derivate, which can drastically improve the transient behaviour of the system, as shown in Figure 8, where controllers are controlling a mass-spring-damper system.

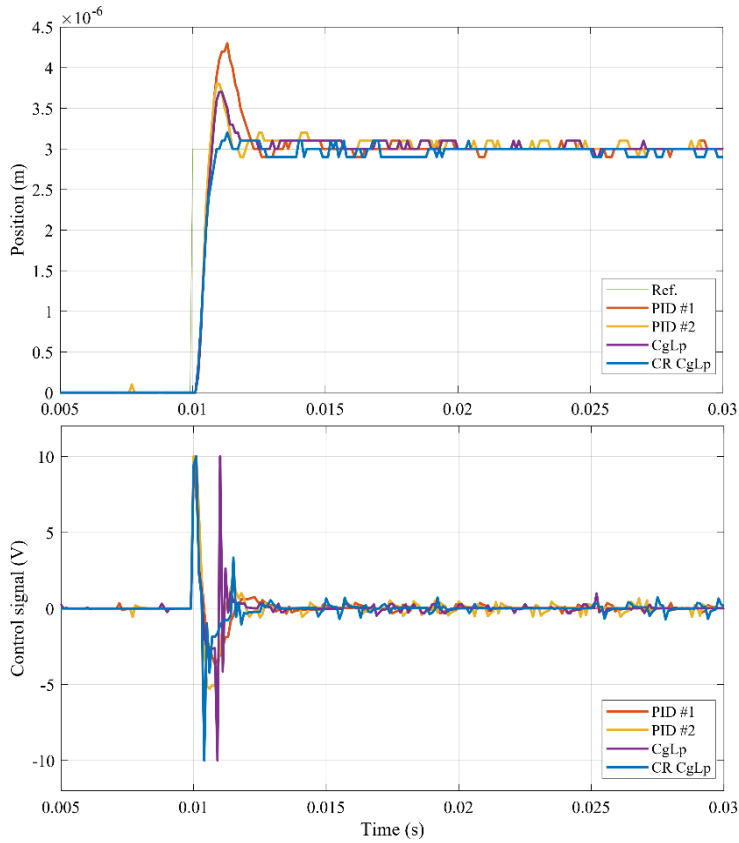

Figure 8: Strep response of Continous Reset CgLp (CR CgLp) compared to a conventional CgLp and two different PIDs.

Furthermore, it is shown that in this architecture, the presence of a lag filter, i.e. $R(s)$, reduces the higher-order harmonics of the controller while keeping the desirable almost first-harmonic intact. Figure 9 shows the comparison of first- and higher-order harmonics of a CR CgLp and a conventional CgLp.

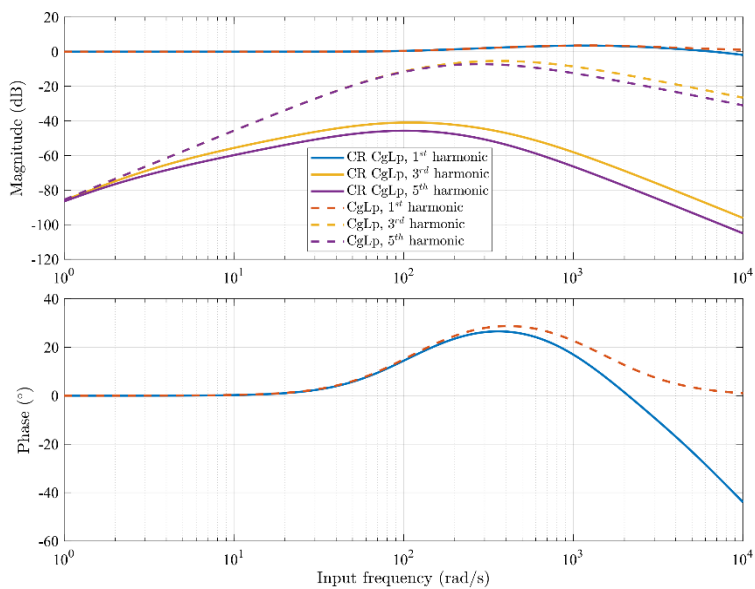

Figure 9: First- and higher-order harmonics of a CR CgLp and a conventional CgLp.

The main drawback of such an architecture is the fact that $L(s)$ will amplify the high-frequency content of noise in closed-loop error signal, which can create excessive unwanted reset instants. Such problems can be solved by utilising the Kalman filter in the architecture shown in Figure 10.

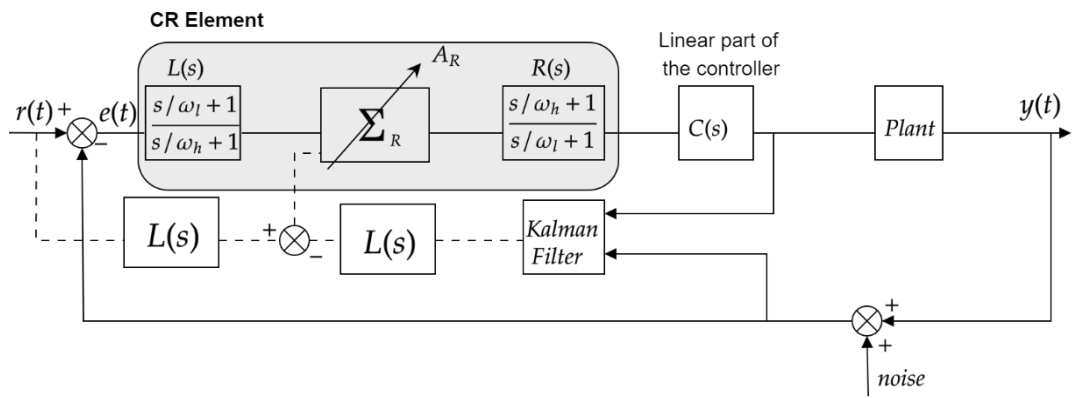

Figure 10: Closed-loop architecture containing Kalman filter to reduce the effect of noise on resetting conditions.

However, in many precision motion applications, the presence of a feed-forward controller will take care of the transient tracking response, and the feedback loop is mainly responsible for disturbance rejection. In such cases, a parallel equivalent of $L(s)$ can be put in parallel to reset element instead of series. This will prevent $L(s)$ from creating of excessive resets while keeping the benefit $R(s)$. Figure 11 shows such configuration.

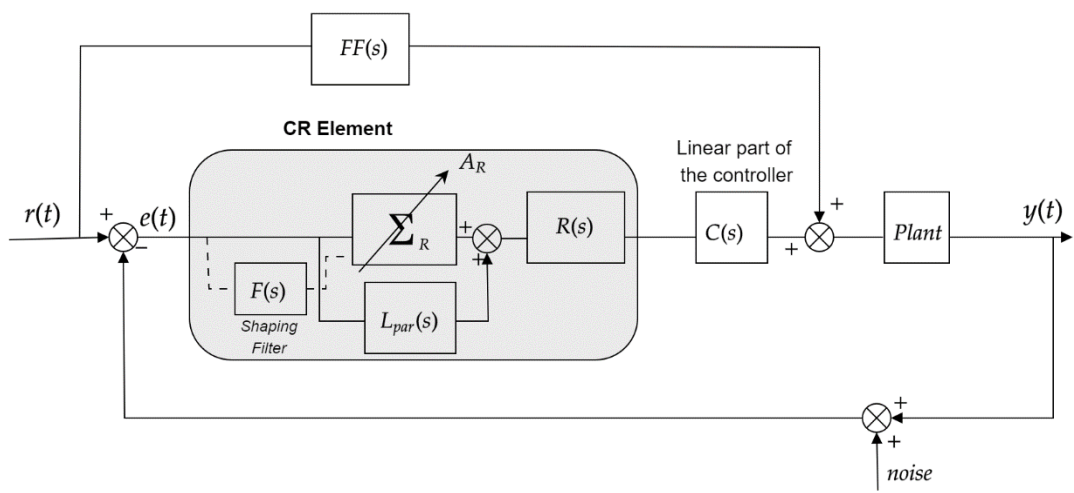

Figure 11: Closed-loop architecture containing parallel CR architecture.

Assuming that $B(s)$ is the base-linear transfer function of the reset element, $L_{par}(s)$ should be chosen in a manner that $L_{par}(s) + B(s) = L(s)B(s)$. The presence of shaping filter on the reset line can shape the nonlinearity based on the priciples discussed in [11, 12].